\documentstyle[12pt]{article}
\begin{document}
\begin{titlepage}
\rightline{\bf UMSNH-PHYS/01-8}

\begin{centering}
 
{\Large\bf Non-minimal coupling for spin $3/2$ fields}\\
\vspace{1.5cm}
Victor M. Villanueva$^{\dag ,}$\footnote{E-mail: 
{\tt vvillanu@ifm1.ifm.umich.mx}},
Juan A. Nieto,$^{\ddag ,}$\footnote{E-mail: {\tt nieto@uas.uasnet.mx}}
and Octavio Obreg\'on,$^{\star ,}$\footnote{E-mail: 
{\tt octavio@ifug3.ugto.mx}}\\
\vspace{0.6cm}
$^{\dag}${\em Instituto de F\'{\i}sica y Matem\'aticas}\\
{\em Universidad Michoacana de San Nicol\'as de Hidalgo}\\
{\em P.O. Box 2-82, Morelia Michoac\'an, M\'exico}\\
\vspace{0.5cm}
$^{\ddag}${\em Facultad de Ciencias F\'{\i}sico-Matem\'aticas}\\
{\em Universidad Aut\'onoma de Sinaloa, 80010, Culiac\'an Sinaloa, M\'exico}\\
\vspace{0.5cm}
$^{\star}${\em Instituto de F\'{\i}sica, Universidad de Guanajuato}\\
{\em P.O. Box E-143, 37150 Le\'on Guanajuato, M\'exico}
\vspace{1cm}
\begin{abstract}

\noindent 
The problem of the electromagnetic coupling for 
spin $3/2$ fields is discussed. Following supergravity and some recent 
works in the field of classical supersymmetric particles, we find 
that the electromagnetic coupling must not obey a minimal coupling in the 
sense that one needs to consider not only the electromagnetic potential but also the coupling of the electromagnetic field strenght. This coupling coincides with the one found by Ferrara {\it et al} by requiring that the gyromagnetic ratio be 2. Coupling with non-Abelian Yang-Mills fields is also discussed.
\end{abstract}

\end{centering} 


\vspace{25pt}


\end{titlepage}

\setcounter{footnote}{0}

\section{Introduction}
\label{Sect1}
The standard treatment of the free masless spin $3/2$ field is achieved by
means of the Rarita-Schwinger (R-S) lagrangian \cite{RS-LS}
\begin{equation}
{\cal L_{RS}} = - {\frac{1}{2}}\epsilon^{\mu\nu\rho\sigma} \bar \Psi_\mu 
\gamma_5 \gamma_\nu \partial_\rho \Psi_\sigma \ \ .
\end{equation}
This lagrangian leads to the field equations   
\begin{equation}
\epsilon^{\mu\nu\rho\sigma} \gamma_5 \gamma_\nu \partial_\rho \Psi_\sigma = 0 \ \ . 
\end{equation}
\noindent

It is however well known that the usual minimal electromagnetic coupling 
prescription for the Dirac field does not work adequatelly for this spin 
$3/2$ field.
In fact if one couples minimally this field with electromagnetism, then several physical inconsistencies arise of which the most remarkable is the appareance of superluminal speed for the particles \cite{VZ}.

By demanding that the scattering amplitudes for arbitrary spin particles should have a good high energy behaviour, Weinberg \cite{Weinberg} showed that the gyromagnetic ratio should be $g \sim 2$. Following a consistent procedure for constructing the lagrangians for higher spin massive particles interacting with the electromagnetic field, Ferrara {\it et al} \cite{FPT} also obtained a gyromagnetic ratio $g = 2$. As a result, their equations of motion contain an extra dipole term that can be implemented at the tree level, thus modifying the usual minimal electromagnetic coupling. A very important feature of this extra dipole term is that, as shown by Ferrara {\it et al}, it avoids the physical inconsistencies for spin $3/2$ particles described in ref. \cite{VZ}.

On the other hand, two of the authors have constructed a theory of the
classical supersymmetric spin $3/2$ particle \cite{NO} in analogy with the classical supersymmetric spin $1/2$ particle formalism developed by Galvao and Teitelboim \cite{GT}. 

In that article, it was shown that the Rarita-Schwinger equations in flat
space-time are the square root of the full linearized Einstein field equations. This is not a consequence of the well known result in canonical 
supergravity \cite{Teitelboim}, where it was shown that the supersymmetry constraint is the square root of the usual Hamiltonian constraint in canonical general relativity. This last procedure involves only some of the dynamical equations, in contrast with the relations found in ref. \cite{NO} which relates the complete set of linearized Einstein field equations and Rarita-Schwinger field equations.

The result of the paper mentioned above shows that the Rarita-Schwinger 
equation is related with linearized gravity as the Dirac equation is related 
to the Klein-Gordon equation. 
Thus, following this analogy and knowing how gravity couples with matter one would expect to be able to find out the way matter couples with the spin $3/2$ field in flat space.

The last point is the main guide for this work, because in principle we can
add a matter tensor on the right side of the linearized Einstein field 
equations and investigate its ``square root" in a similar manner to that 
developed in \cite{NO}. As a result of this procedure a modified R-S equation
will arise. Obviously, this ``square root" must include the terms of 
interaction with the matter fields, and these terms of interaction will give 
us the information for the coupling of spin $3/2$ fields with any kind of 
matter, particularly electromagnetism or Yang-Mills fields.
It is important to mention that this work is a refined version of ref. \cite{NOV}.

This paper is organized as follows: in section 2 we generalize the four indices differential operator representing the linearized general relativity equations \cite{NO} in order to include electromagnetism and non-abelian Yang-Mills fields. 
Based on the particular form of this extra matter term in the linearized
Einstein field equations we search for the particular extra terms in the 
Rarita-Schwinger equation which when squared will produce the desired term.
As a result, we find the interaction for the spin $3/2$ field with 
electromagnetism and Yang-Mills fields. It is to be remarked that this modified R-S equation when squared does not reproduce only the desired extra term in the linearized gravity equations, but there appear extra terms. This is not surprising because the relationship between the R-S equation with interaction and the linearized Einstein field equations is similar to that existing between the Dirac equation with interaction and the Klein-Gordon equation, where the $L S$ coupling term appears. 

Our spin $3/2$ field equation with interaction can be understood as a 
constraint in the classical supersymmetric spin $3/2$ particle formulation
whose squared gives another constraint which is a kind of generalized 
``hamiltonian". In this case the linearized gravity equations with the four 
indices generalized matter tensor.

In some sense our R-S equations can be interpreted as supercharges generating 
the hamiltonian but does not correspond to a canonical formulation. On the other hand, supergravity is the theory that naturally
incorporates in a consistent supersymetrization procedure gravity, spin $3/2$
fields and matter fields. We expect that by linearizing Supergravity we will
be able to reproduce the case without matter \cite{NO}, which will correspond to Supergravity N=1. This is performed in section 3.

In the next two sections we also linearize Supergravity N=2 (section 4) and 
N=4 (section 5). We show that the same kind of interaction found in section 2
for the electromagnetic field and the non-abelian Yang-Mills field respectively
follow. However, in these cases there are correspondingly two and four spin
$3/2$ fields. In each of these linearized Supergravities ( N=2, 4), the
interaction  acts by mixing these R-S fields. The apperance of more spin $3/2$
fields is directly related with the fact that in these last two cases we are 
treating with an enlarged supersymmetry. Supergravity dictates, however, 
esentially the same interaction found by taking the ``square root" of the 
generalized four indices hamiltonian containing  the linearized Einstein field 
equations with matter.

In section 3 we obtain from linearized Supergravity N=1 the Rarita-Schwinger
equations as the square root of the linearized Einstein field equations. 
In section 4 by the same procedure we obtain from linearized Supergravity N=2 
a non-minimal electromagnetic coupling for the spin $3/2$ field. 
It is interesting to mention that this coupling coincides with the one found 
by Ferrara {\it et al}. Nevertheless we must mention that the price of using 
Supergravity N=2 is that we have now two spin $3/2$ fields (the two gravitinos). 
In section 5 we repeat once more the procedure outlined in section 3, but this 
time we apply it on linearized Supergravity N=4 in order to obtain a coupling 
with non-abelian Yang-Mills fields. In this point we apply the formalism over 
four spin $3/2$ fields (the four gravitinos).

\section{Electromagnetic and Yang-Mills generalized energy momentum tensors}
\label{Sect2}

As mentioned in the introduction, we have on one hand the linearized Einstein
field equations  and on the other hand, we have the Rarita-Schwinger equations
as their square root. Thus it is natural to think that we can put an interaction for these Einstein field equations and obtain its square root in order to investigate the possible coupling of the spin $3/2$ field with matter fields.

It has been shown in ref. \cite{NO} that if one associates to the R-S
equation the classical constraint 
\begin{equation}
{\cal S}^{\alpha\beta} \equiv \epsilon^{\alpha\beta\rho\sigma} \theta_\rho P_\sigma = 0 \ \ ,
\label{constraint}
\end{equation}
\noindent
then one has
\begin{equation}
\{ {\cal S}^a_\mu, {\cal S}^\beta_\nu \}   = {\cal H}^{\alpha\beta}_{\mu\nu}  \ \ ,
\end{equation}
\noindent
where 
\begin{equation}
{\cal H}^{\alpha\beta}_{\mu\nu} =  \epsilon^{\alpha~\rho\sigma}_{~\mu} \epsilon^{\beta~\lambda\gamma}_{~\nu} \eta_{\rho\lambda} P_\sigma P_\gamma 
\ \ , 
\label{hamiltonian}
\end{equation}
\noindent
is the ``Hamiltonian'' operator that acts over $h_{\alpha\beta}$ in standard
linearized gravity, {\it i. e.}  
\begin{equation}
{\cal H}^{\alpha\beta}_{\mu\nu} h_{\alpha\beta} = 0 \ \, . 
\end{equation}
\noindent

We first note that in Eq. (\ref{hamiltonian}) the momenta $P_\sigma$ appear quadratically. 
Now we want to introduce the potential term in $ {\cal H}^{\alpha\beta}_{\mu\nu}$,
obviously, this must also be a four indices tensor $ {\cal T}^{\alpha\beta}_{\mu\nu}$,
it should also have units of energy (same as $P_\sigma^2$) and it should be
possible to take its square root in terms of the fields characterizing the matter under consideration.

In particular for the electromagnetic case, the most natural ``potential" ought be constructed as some square of the field $F_{\mu\nu}$. The mathematical 
structure of Eq. (\ref{hamiltonian}) suggests us to accompany the $F^2$ term by two Levi-Civitta
tensors, {\it i. e.}

\begin{equation}
{\cal T}^{\alpha\beta}_{\mu\nu} \sim  
\epsilon^{\alpha~\rho\sigma}_{~\mu} \epsilon^{\beta~\lambda\gamma}_{~\nu} 
\big( F_{\rho\sigma}F_{\lambda\gamma} + \Lambda \tilde F_{\rho\sigma}F_{\lambda\gamma}
+ \kappa \tilde F_{\rho\sigma} \tilde F_{\lambda\gamma} \big)
 \ \, . 
\label{potential}
\end{equation}
\noindent
where $\Lambda ~{\rm and } ~\kappa$ are constants and $\tilde F_{\rho\sigma}$ is the dual of $ F_{\rho\sigma}$.

In order to achieve the square root of the ``hamiltonian" (\ref{hamiltonian}) plus the
``potential" (\ref{potential}) we first notice that the desired interaction term in the
R-S constraint (\ref{constraint}), must alone give us when squared, the ``potential" term
$ {\cal T}^{\alpha\beta}_{\mu\nu}$. Thus, the most general construction one can 
propose is a linear combination of $F_{\rho\sigma}$ plus its dual. 
Nevertheless Dirac matrices must be introduced in the linear combination of
the fields $F_{\rho\sigma}$ and its dual since we are applying these constraints 
over four components spinors.

Then the desired interaction term in the R-S constraint is of the form
\begin{equation}
F^{\mu\nu} + \kappa \epsilon^{\mu\nu\rho\sigma} F_{\rho\sigma} \ \ ,
\end{equation}
\noindent
where now $\kappa$ is a Dirac matrix or a product of Dirac matrices.

A natural generalization of this result to the case of non-abelian Yang-Mills
would be the tensor
\begin{equation}
 T^{\alpha\beta}_{\mu\nu} \sim  
\epsilon_{~ \mu}^{\alpha ~ \rho\sigma} \epsilon_{~ \nu}^{\beta ~ \lambda\gamma} 
\big(  F^a_{\rho\sigma} F^a_{\lambda\gamma} + \Lambda \tilde F^a_{\rho\sigma} F^a_{\lambda\gamma}
+ \kappa \tilde F^a_{\rho\sigma} \tilde F^a_{\lambda\gamma} \big) \ \ ,
\end{equation}
\noindent
where $F^a_{\rho\sigma}$ is the Yang-Mills field tensor, and the corresponding interaction will be also of the form (\ref{constraint}) with appropriate indices.

\section{Free massless spin $3/2$ field}
\label{Sect3}

In this section we will review the calculations of the main result of ref. \cite{NO}. The reasons of doing so are just pedagogical.

The langrangian for {\it Supergravity N=1} \cite{Niew} is given by
\begin{equation}
{\cal L} = -{ e \over 2}{\cal R} -{ e \over 2}{\bar\Psi_\mu}\Gamma^{\mu\rho\sigma}D_\rho \Psi_\sigma \ \ , 
\end{equation}
\noindent
where $e$ is the determinant of the tetrad, ${\cal R}$ is the generalized curvature, 
$\Psi_\mu$ is the gravitino field (spin $3/2$), and 
$\Gamma^{\mu\rho\sigma}=\epsilon^{\mu\nu\rho\sigma}\gamma_5\gamma_\nu$. 

Once more the equations of the motion for the gravitino field are found to be 
\begin{equation}
\epsilon^{\mu\nu\rho\sigma} \gamma_5 \gamma_\nu \partial_\rho \Psi_\sigma = 0 \ \ ,
\label{graveq}
\end{equation}
\noindent
we can associate to Eq. (\ref{graveq}) the classical constraint
\begin{equation}
{\cal S}^{\mu\nu} \equiv \epsilon^{\mu\nu\rho\sigma} \theta_\rho P_\sigma = 0 \ \ ,  
\end{equation}
\noindent
where $\theta_\rho = {1\over \sqrt 2} \gamma_5 \gamma_\rho $,  
$\theta_5 = {1\over \sqrt 2} \gamma_5 $ are the classical 
limit of the gamma matrices of Dirac and the operator $P_\sigma = - i \partial_\sigma$ ($\hbar = 1$).

Considering that the only nonvanishing Poisson brackets [5] between these variables are
\begin{eqnarray}
\big\{ \theta_\mu , \theta_\nu \big\} & = &  \eta_{\mu\nu} \ \ , \\
\big\{ \theta_5 , \theta_5 \big\}     & = & 1 \ \ , \\
\big\{ x_\mu , P_\nu \big\}           & = & i \eta_{\mu\nu} \ \ , \\
\nonumber
\end{eqnarray}
\noindent
we get the algebra
\begin{equation}
\{ {\cal S}^a_\mu, {\cal S}^\beta_\nu \}   = {\cal H}^{\alpha\beta}_{\mu\nu} \ \ ,
\end{equation} 
\noindent
where 
\begin{equation}
{\cal H}^{\alpha\beta}_{\mu\nu} =  \epsilon^{\alpha~\rho\sigma}_{~\mu} \epsilon^{\beta~\lambda\gamma}_{~\nu} \eta_{\rho\lambda} P_\sigma P_\gamma \ \ ,
\end{equation} 
\noindent
is the ``Hamiltonian'' operator that acts over $h_{\alpha\beta}$ in standard
linearized gravity, {\it i. e.}  
\begin{equation}
{\cal H}^{\alpha\beta}_{\mu\nu} h_{\alpha\beta} = 0  \ \ .
\end{equation} 
\noindent

As claimed in the preceding section, the Rarita-Schwinger equations turn out 
be the square root of the linearized Einstein field equations. Obviously, we 
have no contribution of any other matter field, since we have no interaction
at all. Nevertheless this result will be the guide to investigate the coupling
of Rarita-Schwinger fields, in principle, with any kind of matter as will be developed in the next sections.

\section{Electromagnetic interaction of spin $3/2$ fields}
\label{section4}
In order to investigate the electromagnetic coupling of spin $3/2$ fields we use the resource of {\it Supergravity N=2} \cite{Niew}, since it naturally incorporates the graviton, the electromagnetic field and two gravitinos.

The lagrangian for Supergravity $N = 2$ is 
\begin{eqnarray}
{\cal L}  & = &  -{ e \over 2}{\cal R} -{e\over 2}{\bar\Psi^i_\mu}\Gamma^{\mu\rho\sigma}D_\rho{\Psi^i_\sigma}
-{ e \over 4}F_{\alpha\beta} F^{\alpha\beta}  \\ 
& + & {\kappa \over 4\sqrt2}{\bar\Psi^i_\mu}\big[e(F^{\mu\nu} + \hat F^{\mu\nu}) 
+ {\scriptstyle{1\over 2}}\gamma_5({\tilde F^{\mu\nu}} + {\skew4 \tilde {\hat F}^{\mu\nu}})\big]
{\bar\Psi^j_\nu}\epsilon^{ij} \ \ , 
\nonumber
\end{eqnarray}
\noindent
where $e$ is the determinant of the metric, $\cal R$ is the curvature, 
$\Gamma^{\mu\rho\sigma} = \epsilon^{\mu\lambda\rho\sigma}\gamma_5\gamma_\lambda$,
$D_\rho = \partial_\rho + {1\over 2}\omega_\rho^{mn}\sigma_{mn}$, is the derivative including the spin connection, $\omega_\rho^{mn}$, $\sigma_{mn} = {1\over 4}[\gamma_m,\gamma_n]$, 
$\tilde F^{\mu\nu} = \epsilon^{\mu\nu\alpha\beta} F_{\alpha\beta}$,
and
\begin{equation}
\hat F_{\mu\nu} = \partial_\mu A_\nu - \partial_\nu A_\mu - {\kappa \over 2\sqrt2}[\Psi^i_\mu \Psi^j_\nu - \Psi^i_\nu \Psi^j_\mu]\epsilon^{ij} \ \ ,
\end{equation}
\noindent
is the supercovariant curl.

By performing variations with respect to $\bar\Psi^l_\alpha$ in the action, we obtain the equations of the motion for the gravitinos
\begin{eqnarray}
&-&{e \over 2}\Gamma^{\alpha\rho\beta}D_\rho{\Psi^l_\beta} 
+ {\kappa \over 4\sqrt2} \big\{ \big[ e \big( 2F^{\alpha\beta} - {\kappa \over 2\sqrt2} (\bar\Psi^{i\alpha} \Psi^{j\beta} - \bar\Psi^{i\beta} \Psi^{j\alpha}) \epsilon^{ij} \big) \nonumber \\
&+& {\scriptstyle{1\over 2}}\gamma_5 \epsilon^{\alpha\beta\rho\sigma}\big( 2 F_{\rho\sigma} - {\kappa \over 2\sqrt2}(\bar\Psi^i_\rho \Psi^j_\sigma - \bar\Psi^i_\sigma \Psi^j_\rho) \epsilon^{ij} \big) \big] \Psi^k_\beta\epsilon^{lk} \\
&-& {\kappa \over 2\sqrt2}\bar\Psi^{k\beta} \big[ e(\delta^{\alpha\beta} \Psi^{j\gamma} - \delta^{\alpha\gamma} \Psi^{j\beta} ) + {\scriptstyle{1\over 2}}\gamma_5 \epsilon^{\beta\gamma\rho\sigma}(\delta^\alpha_\rho \Psi^j_\sigma - \delta^\alpha_\sigma \Psi^j_\rho) \big] 
\Psi^h_\gamma\epsilon^{lj}\epsilon^{hk} \big \} = 0 \ \ . \nonumber \\ 
\nonumber
\label{eqgrav}
\end{eqnarray}
\noindent

Linearizing the above equation by eliminating gravitational interactions
and neglecting terms of the order $\Psi^3$, the field equations reduce to
\begin{equation}
\epsilon^{\alpha\mu\nu\beta}\gamma_5\gamma_\mu \partial_\nu \Psi^i_\beta - {\kappa \epsilon^{ij} \over {\sqrt 2}} \big[ 
F^{\alpha\beta} + {\scriptstyle{1\over 2}}\gamma_5 \epsilon^{\alpha\beta\rho\sigma} F_{\rho\sigma} \big]
\Psi^j_\beta  = 0 \ \ ,
\end{equation}
\noindent
notice that the term in squared brackets is 
$F^{+\alpha\beta} = F^{\alpha\beta} + {\scriptstyle{1\over 2}}\gamma_5 \epsilon^{\alpha\beta\rho\sigma} F_{\rho\sigma} $
precisely the dipole term found by Ferrara {\it et al} \cite{FPT}. 
In their article they have shown that this term cancells divergences and avoids superluminal velocities in systems of spin $3/2$ particles.

Equation (\ref{eqgrav}) can be shown to be the generalized Rarita-Schwinger equation
\begin{equation}
\epsilon^{\alpha\beta\mu\nu} \big[ \delta_{ij}\theta_\mu P_\nu + i \epsilon_{ij}{\cal F}_{\mu\nu} \big] \Psi^j_\beta = 0 \ \ , 
\label{genrs}
\end{equation}
\noindent
where ${\cal F}_{\mu\nu} = {\kappa \over \sqrt 8} [ {1 \over \sqrt 8} \tilde F_{\mu\nu} + \theta_5 F_{\mu\nu} ] $ 
is a ``rotation" of the dipole term $F^{+\mu\nu}$.
\noindent 
Obviously, the solutions of these equations are the cuasiclasical ones, and it 
becomes clear that as a consequence of Supergravity, the solutions of our equation 
must not have physical inconsistencies, such as superluminal motion \cite{VZ}. 
Thus, we can associate to Eq. (\ref{genrs}) the constraint
\begin{equation}
{\cal S}^{\alpha\beta}_{ij} = \epsilon^{\alpha\beta\mu\nu} \big[ \delta_{ij}\theta_\mu P_\nu + i \epsilon_{ij}{\cal F}_{\mu\nu} \big] = 0 \ \ , 
\end{equation}
\noindent
and by using the Poisson brackets of Section 3, we get the algebra
\begin{eqnarray}
\big\{ {\cal S}^{\alpha ~ij}_{~\mu} , {\cal S}^{\beta ~kl}_{~\nu} \big\} 
& = &
\epsilon^{\alpha~\rho\sigma}_{~\mu} \epsilon^{\beta~\lambda\gamma}_{~\nu} 
\big[ \eta_{\rho\lambda} P_\sigma P_\gamma \delta^{ij} \delta^{lk}  
-  {\kappa^2 \over 8} F_{\rho\sigma}F_{\lambda\gamma} \epsilon^{ij} \epsilon^{lk}  \nonumber \\
& + & i \big( \theta_\rho {\cal F}_{\lambda\gamma , \sigma} \delta^{ij} \epsilon^{lk} +  
\theta_\lambda {\cal F}_{\rho\sigma , \gamma}  \epsilon^{ij} \delta^{lk} \big) \big] \ \ . \\
\nonumber
\end{eqnarray}

The first term in the last equation is the Hamiltonian for linearized gravity
discussed before
\begin{equation}
\epsilon^{\alpha~\rho\sigma}_{~\mu} \epsilon^{\beta~\lambda\gamma}_{~\nu} 
\eta_{\rho\lambda} P_\sigma P_\gamma \delta^{ij} \delta^{lk} =  {\cal H}^{\alpha\beta}_{\mu\nu} \delta^{ij} \delta^{lk} \ \ ,
\end{equation}
\noindent
the second of these terms is the generalized energy momentum for the 
electromagnetic field announced in Eq. (\ref{potential}), that is
\begin{equation}
{\kappa^2 \over 8} \epsilon^{\alpha~\rho\sigma}_{~\mu} \epsilon^{\beta~\lambda\gamma}_{~\nu} F_{\rho\sigma}F_{\lambda\gamma} \epsilon^{ij} \epsilon^{lk} = {\cal T}^{\alpha\beta}_{\mu\nu} \epsilon^{ij} \epsilon^{lk} \ \ ,
\end{equation}
\noindent
where this tensor has the form
\begin{eqnarray}
{\cal T}^{\alpha\beta}_{\mu\nu} & = & {\kappa^2 \over 8} 
\big[ 2 \eta_{\mu\nu} \eta^{\alpha\beta} F_{\rho\sigma}F^{\rho\sigma} 
-2 \delta^\alpha_\nu \delta^\beta_\mu F_{\rho\sigma}F^{\rho\sigma} 
-2 F^{\alpha~}_{~\mu}F_\nu^{~\beta} \nonumber \\
& + & 4 \eta^{\alpha\beta} F_{\mu\rho} F^{\rho~}_{~\nu}
+ 4 \eta_{\mu\nu} F^{\alpha\rho} F_\rho^{~\beta}
- 4 \delta^\alpha_\nu F_{\mu\rho} F^{\rho\beta} 
+ 4 \delta^\beta_\mu F^{\alpha\rho} F_{\nu\rho}\big] \ \ . \\
\nonumber
\end{eqnarray}

The third term gives an electromagnetic interaction for the gravitinos, this
term is 
\begin{equation}
\epsilon^{\alpha~\rho\sigma}_{~\mu} \epsilon^{\beta~\lambda\gamma}_{~\nu} 
\big( \theta_\rho {\cal F}_{\lambda\gamma , \sigma} \delta^{ij} \epsilon^{lk} + \theta_\lambda {\cal F}_{\rho\sigma , \gamma}  \epsilon^{ij} \delta^{lk} \big) \ \ .
\end{equation}
\noindent
This term contains a coupling between the gradient of the  electromagnetic 
tensor field $ {\cal F}_{\rho\sigma , \gamma} $ and the spin tensor 
$S_{\mu\nu} = i \theta_\mu \theta_\nu $. 

It is interesting to comment  that $ {\cal T}^{\alpha\beta}_{\mu\nu} $ may be 
understood only as part of a total energy momentum tensor that contains now also $ {\cal F}^{\alpha\beta}_{\mu\nu}$. In this way, our model predicts a 
coupling between gravity, the gradient of the electromagnetic field tensor 
and the spin tensor, thus the nonminimal coupling for spin $3/2$ 
fields is now given by ($\ref{genrs}$).

\section{Yang-Mills field interaction of spin $3/2$ fields}
\label{section5}
Now we are in a position to explore the possibility that a Yang-Mills field be 
coupled to a Rarita-Schwinger field. The simplest supergravity model that 
involves a non-Abelian Yang-Mills field is {\it Supergravity N=4} \cite{Niew}.

In the philosophy of the preceding calculations, we can associate a classical
constraint to the field equation for the gravitinos. It turns out to be
\begin{eqnarray}
{\cal S}^{\alpha\beta}_{ij} & = & {i \over \sqrt 2}\epsilon^{\alpha\beta}_{~~\mu\nu} \delta_{ij}\theta^\mu P^\nu - {\kappa \over 2 \sqrt 2} \big( {\cal F}^{\mu\nu}_{(ij)} - {i \over \sqrt 2}\theta_5 \epsilon^{\alpha\beta}_{~~\rho\sigma}{\cal F}^{\rho\sigma}_{(ij)} \big)  \nonumber \\
& - & {{\sqrt 2} \delta_{ij} \over 4 \kappa} \big( e_A + {\sqrt 2} i e_B \theta_5 \big) \sigma^{\alpha\beta} \ \ , \\
\nonumber
\end{eqnarray} 
\noindent
where now the contribution of the non-Abelian field is given 
\begin{equation}
{\cal F}^{\rho\sigma}_{(ij)} = \alpha^k_{(ij)} A^{\rho\sigma}_k + {\sqrt 2} i
\theta_5  \beta^k_{(ij)} B^{\rho\sigma}_k \ \ ,
\label{nafield}
\end{equation}
\noindent
and 
\begin{eqnarray}
A_{\rho\sigma}^k  & = & \partial_\rho A^k_\sigma - \partial_\sigma A^k_\rho + e_A \epsilon^{ijk} A^i_\rho A^j_\sigma \ , \\
B_{\rho\sigma}^k  & = & \partial_\rho B^k_\sigma - \partial_\sigma B^k_\rho + e_B \epsilon^{ijk} B^i_\rho B^j_\sigma \ , \\
\nonumber
\end{eqnarray}
\noindent
are the non-Abelian Yang-Mills fields. In Eq. (\ref{nafield}) the alpha's and beta's are matrices that generate the $SU(2)\times SU(2)$ symmetry of supergravity N=4.

By using the Poisson brackets of the preceding section, we get the algebra
\begin{eqnarray}
\big\{ {\cal S}^\alpha_{~\mu ij} , {\cal S}^\beta_{~\nu kl} \big\} 
& = &  - {1 \over 2}\epsilon^{\alpha~\rho\sigma}_{~\mu} \epsilon^{\beta~\lambda\gamma}_{~\nu} \big[ \eta_{\rho\lambda} P_\sigma P_\gamma \delta_{ij} \delta_{lk}  + {\kappa^2 \over 4}{\cal F}_{\rho\sigma (ij)}{\cal F}_{\lambda\gamma (kl)} \big] \nonumber \\
& - & {i \kappa \over 4} 
\big[\epsilon^{\alpha~\rho\sigma}_{~\mu}\delta_{ij} \theta_\rho \big( {\cal F}^\beta_{~\nu,\sigma (kl)} - {i \over \sqrt 2} \theta_5
\epsilon^{\beta~\lambda\gamma}_{~\nu}{\cal F}_{\lambda\gamma,\sigma (kl)} \big) \nonumber \\
& - &   \epsilon^{\beta~\lambda\gamma}_{~\nu }\delta_{kl} \theta_\lambda \big( {\cal F}^\alpha_{~\mu,\gamma (ij)} - {i \over \sqrt 2} \theta_5
\epsilon^{\alpha~\rho\sigma}_{~\mu} {\cal F}_{\rho\sigma,\gamma (ij)} \big) 
\big] \\
& + & {e_B\over 8} \big[ \epsilon^{\alpha~\rho\sigma}_{~\mu} \sigma^\beta_{~\nu}
\delta_{kl} {\cal F}_{\rho\sigma (ij)} + \epsilon^{\beta~\lambda\gamma}_{~\nu} \sigma^\alpha_{~\mu} \delta_{ij}{\cal F}_{\lambda\gamma (kl)} \big] 
\nonumber \\
& + & {1 \over 8 \kappa^2} \delta_{ij} \delta_{kl} \bigg[ 
e_B^2 \big( \theta_\nu \theta^\beta - \theta^\beta \theta_\nu \big) \big( \theta_\mu \theta^\alpha - \theta^\alpha \theta_\mu \big) \nonumber \\
& + & \big( {1\over 2}(e_A^2 - e_B^2) + 2{\sqrt 2}i e_A e_B \theta_5 \big) \big(\eta^{\alpha\beta}\theta_\mu \theta_\nu - \eta_{\mu\nu}\theta^\beta \theta^\alpha + \delta^\alpha_\nu \theta_\mu \theta^\beta - \delta^\beta_\mu \theta_\nu \theta^\alpha \big) \bigg] \ \ . \nonumber \\
\nonumber
\end{eqnarray}
Once more, we can identify terms and the first of them is exactly the same 
tensor found in ref. \cite{NO}, it give us the linearized operator for the Einstein field equations. That is 
\begin{equation}
\epsilon^{\alpha~\rho\sigma}_{~\mu} \epsilon^{\beta~\lambda\gamma}_{~\nu} 
\eta_{\rho\lambda} P_\sigma P_\gamma \delta_{ij} \delta_{lk} =  {\cal H}^{\alpha\beta}_{\mu\nu} \delta_{ij} \delta_{lk} \ \ , 
\end{equation}

The second term is the analogous of the generalized electromagnetic field 
energy momentum tensor, that in this case is a non-Abelian gauge field energy 
momentum tensor.
\begin{equation}
\epsilon^{\alpha~\rho\sigma}_{~\mu} \epsilon^{\beta~\lambda\gamma}_{~\nu}  
{\cal F}_{\rho\sigma (ij)}{\cal F}_{\lambda\gamma (kl)} =  {\cal T}^{\alpha\beta}_{\mu\nu(ijkl)} \ \ .
\end{equation}
\noindent 
The following terms contain  couplings of the spin tensor 
$S_{\mu\nu}= i \theta_\mu \theta_\nu$ and to the gradient of the non-Abelian 
field tensor ${\cal F}_{\rho\sigma (ij)}$. 

Thus we have obtained a nonminimal coupling once more for spin $3/2$ particles. Moreover the coupling obtained generalizes that of the dipole term found by Ferrara {\it et al}.

\section{Conclusions}
\label{Sect6}

We have discussed the problem of the non-minimal coupling for the Rarita-Schwinger fields, and the attempts of Weinberg and Ferrara {\it et al} to solve the problem by demanding $g = 2$ for arbitrary spin particles. As a result, they have obtained an extra dipole term in the equations for these fields. This dipole term avoids the bad energy behavior of the particles and the physical inconsistencies discussed in ref. \cite{VZ}. 

By using the fact that the Rarita-Schwinger field equations are the square 
root of the linearized Einstein field equations as a guide, we have implemented energy momentum tensors for electromagnetic and non-abelian Yang-Mills fields.
After that, linearized supergravity N=2 was used as a tool that provided us 
with field equations for the Rarita-Schwinger fields. Surprisingly, by squaring the constraints associated to these equations we obtained an energy momentum equal to that announced in section 2. Besides we obtained for the fields a non-minimal coupling consisting of a ``rotation'' of the dipole term found by Ferrara {\it et al}.

A similar analysis for non-abelian Yang-Mills fields was developed by using 
supergravity N=4. Thus obtaining a similar energy momentum tensor to that 
claimed in section 2. The coupling term in this case has a similar structure
to that of the electromagnetic case. It consist of terms of the type
$$Field + {1\over 2} \gamma_5 Dual~ Field. $$
\noindent
It is interestring to mention that a term of similar structure was implemented
by Cucchieri, Porrati and Deser \cite{CPD} by studying the gravitational coupling of 
higher spin fields, where the {\it Field} of the above expression is the 
Riemman tensor.

Further developments of this formalism are being considered. For instance the massive spin
$3/2$ particle interacting with electromagnetic and Yang-Mills fields \cite{DPW}. Another interesting issue is the quantization and possible phenomenological implications of the theory
\cite{Pasca}.

\section{Acknowledgments}

This work has been partially supported by CONACyT through the projects:
28454-E, I-32819-E, University of Michoac\'an through the project CIC 4.14 and University of Sinaloa.

\end{document}